\documentclass[pre,amsmath,twocolumn,showpacs]{revtex4-1}
\usepackage{graphicx}
\usepackage{color}
\usepackage{latexsym}
\usepackage{amssymb}

\begin{document}

\title{Proof of the Finite-Time Thermodynamic Uncertainty Relation for Steady-State Currents}

\author{Jordan M. Horowitz}
\author{Todd R. Gingrich}
\affiliation{Physics of Living Systems Group, Department of Physics, Massachusetts Institute of Technology, 400 Technology Square, Cambridge, MA 02139}

\date{\today}

\begin{abstract}   
The thermodynamic uncertainty relation offers a universal energetic constraint on the relative magnitude of current fluctuations in nonequilibrium steady states.  However, it has only been derived for long observation times.  Here, we prove a recently conjectured finite-time thermodynamic uncertainty relation for steady-state current fluctuations.  Our proof is based on a quadratic bound to the large deviation rate function for currents in the limit of a large ensemble of many copies.
\end{abstract}

\maketitle 
 
\section{Introduction}
The thermodynamic uncertainty relation offers a fundamental bound on the current fluctuations in nonequilibrium steady states~\cite{Barato2015,Pietzonka2016Universal,Gingrich2016,Gingrich2017}.
Roughly speaking, it states that small fluctuations come at the cost of more dissipation.
This relation, and its cousins~\cite{Polettini2016,Maes2017}, allows one to constrain thermodynamic forces in enzymatic catalysis~\cite{Pietzonka2016Affinity,Barato2015Fano}, bound the power fluctuations in mesoscopic machines~\cite{Pietzonka2016MolecularMotors, Pietzonka2017universal}, limit the energetic cost of sensing~\cite{Barat2015Dispersion}, and has been adapted to Brownian motion~\cite{Hyeon2017,Gingrich2017}, nonequilibrium self-assembly~\cite{Nguyen2016}, active matter~\cite{Falasco2016}, equilibrium order parameter fluctuations~\cite{Guioth2016}, activity fluctuations~\cite{Garrahan2017} and first-passage-time fluctuations~\cite{Garrahan2017,Gingrich2017FPT}.
Its derivation relies on bounding the likelihood of rare current fluctuations in the long-time limit, using the tools of large deviation theory~\cite{Touchette2009}.
As such, the predictions are proved to be valid only for long observation times~\cite{Gingrich2016}.

Recently though, Pietzonka, Ritort, and Seifert conjectured a steady-state-currents uncertainty relation valid for finite observation times, based on extensive numerical and experimental evidence~\cite{Pietzonka2017}.
This finite-time uncertainty relation stipulates that in the steady state any current $J_T$ integrated up to time $T$ will have a variance ${\rm Var}(J_T)$ and mean $\langle J_T\rangle$ constrained by the total steady-state entropy production $\langle \Sigma_T \rangle$ accumulated by $T$ as
\begin{equation}\label{eq:TUR}
{\rm Var}(J_T)/\langle J_T\rangle^2\ge 2/\langle \Sigma_T\rangle ,
\end{equation}
in units where Boltzmann's constant is set to $k_{\rm B}=1$.
For the special case of the fluctuating entropy production $\Sigma_T$, the uncertainty relation simplifies to
\begin{equation}
{\rm Var}(\Sigma_T)\ge 2\langle \Sigma_T\rangle,
\end{equation}
which has been derived directly from the structure of entropy-production fluctuations for nonequilibrium systems modeled as diffusion processes~\cite{Pigolotti2017}.
In this communication, we provide a proof of \eqref{eq:TUR} using the tools of large deviation theory in a manner akin to the proof of the original long-time uncertainty relation~\cite{Gingrich2016}.
This proof puts the finite-time uncertainty relation on firm footing, justifying its use in analyzing even short-time experimental data.

\section{Setup}

We have in mind a nonequilibrium system with states $x=1,\cdots, M$.
Transitions between states, say from $z$ to $y$, are modeled as a continuous-time Markov jump process with rates $r_{yz}$.
We assume that the matrix of transition rates is irreducible, so that there is a unique steady-state distribution $\pi_y$ with steady-state current $j_{yz}^\pi=r_{yz}\pi_z-r_{zy}\pi_y$.
In addition, we assume that the transition rates are thermodynamically consistent, so that every ratio of transition rates can be related to a thermodynamic force $F_{yz}=\ln(r_{yz}\pi_z/r_{zy}\pi_y)$, which measures the total entropy production -- environmental entropy flow and change in system Shannon entropy -- along that transition~\cite{Seifert2012}.

Now, as we track a stochastic realization of our system evolving over a finite time interval $t\in[0,T)$, $x(t)$, there will be a fluctuating instantaneous current counting every time $t_k$ the system jumps:
\begin{equation}
\label{eq:current}
j_{yz}(t)=\sum_k\delta(t-t_k) \left(\delta_{x(t_k^+),y}\delta_{x(t_k^-),z}-\delta_{x(t_k^+),z}\delta_{x(t_k^-),y}\right),
\end{equation}
with $x(t_k^\pm)$ being the state of the system just before and after a jump.
Our interest, though, is in  integrated generalized currents, which are obtained by weighing each mesoscopic jump by a factor $d_{yz}(t)=-d_{zy}(t)$ and summing them up: 
\begin{align}\label{eq:Jd}
J_T=\int_0^Tds\, \sum_{y<z}d_{yz}(s)j_{yz}(s).
\end{align}
For example, the entropy production is a generalized current with $d_{yz}=F_{yz}$,
\begin{equation}
\Sigma_T=\int_0^Tds\, \sum_{y<z}F_{yz}j_{yz}(s)
\end{equation}
whose steady-state average 
\begin{equation}
\langle \Sigma_T \rangle=T\sum_{y<z}F_{yz}j_{yz}^\pi\equiv T\Sigma^\pi
\end{equation}
characterizes the irreversibility of the nonequilibrium steady state.
Our goal now is to constrain the fluctuations in $J_T$ by bounding its large deviation rate function using $\langle \Sigma_T \rangle =T\Sigma^\pi$, which will lead to \eqref{eq:TUR}.

\section{Large deviations for large ensembles}

Imagine now not just one instance of our system hopping among its states, but an ensemble of $N\gg1$ independent copies -- labeled $x^\alpha(t)$, $\alpha=1,\dots,N$ -- with initial conditions sampled from the steady-state distribution $\pi$.
Then in any given moment we could obtain an empirical estimate of the density to be in mesostate $y$ at time $t$ by measuring the instantaneous fraction of copies in state $y$:
\begin{equation}
\rho_y(t)=\frac{1}{N}\sum_{\alpha=1}^N \delta_{x^\alpha(t),y}.
\end{equation}
We could additionally estimate the current by counting the total net number of jumps along any link as
\begin{equation}
\phi_{yz}(t)=\frac{1}{N}\sum_{\alpha=1}^Nj^\alpha_{yz}(t),
\end{equation}
with $j^\alpha_{yz}(t)$ the instantaneous current of copy $\alpha$ (cf.~\eqref{eq:current}).
Indeed, the law of the large numbers guarantees that both empirical measures converge to their expected values as $N\to\infty$.
However, we can also quantify their fluctuations through a large deviation principle. 
As demonstrated in \cite{Maes2008On}, the probability to see a fluctuation is exponentially suppressed for large $N$ as
\begin{equation}
{\mathcal P}[\rho(t),\phi(t)]\asymp e^{-NI[\rho(t),\phi(t)]},
\end{equation}
where $\asymp$ denotes asymptotic logarithmic equivalence~\cite{Touchette2009}, and the large deviation rate function is
\begin{equation}
I[\rho(t),\phi(t)]=\int_0^Tds\, {\mathcal I}(\rho(s),\phi(s))-S(\rho(0)||\pi).
\end{equation}
The second term is the relative entropy between the initial fluctuating density $\rho(0)$ and the steady state $\pi$, $S(\rho(0)||\pi)=\sum_x\rho_x(0)\ln(\rho_x(0)/\pi_x)$.
The first term can be put in the form~\cite{Bertini2015Flows,Bertini2015Large}
\begin{equation}\label{eq:Iinstant}
{\mathcal I}(\rho(t),\phi(t))=\sum_{y<z}\Psi(\phi_{yz}(t),j_{yz}^\rho(t),a^\rho_{yz}(t))
\end{equation}
with
\begin{equation}
\begin{split}
\Psi(j,{\bar j},a)=j&\left(\mathrm{arcsinh}\frac{j}{a}-\mathrm{arcsinh}\frac{{\bar j}}{a}\right)\\
&-\left(\sqrt{a^2+j^2}-\sqrt{a^2+{\bar j}^2}\right),
\end{split}
\end{equation}
$j^\rho_{yz}(t)=r_{yz}\rho_z(t)-r_{zy}\rho_y(t)$ the expected current for density $\rho$, and $a^\rho_{yz}(t)=2\sqrt{\rho_y(t)\rho_z(t)r_{yz}r_{zy}}$.
The expression for $I$ only applies for fluctuations that conserve probability, ${\dot \rho}_y(t)=\sum_{z\neq y}\phi_{yz}(t)$ with a normalized density $\sum_y\rho_y(t)=1$; otherwise, $I$ is infinity.

Within this framework, the fluctuations in the generalized current are simply due to the sum over the fluctuations of each member 
\begin{align}
\Phi_{\rm d}&=\sum_{\alpha=1}^N\left(\int_0^T ds\, \sum_{y<z}d_{yz}(s)j^\alpha_{yz}(s)\right)\\
&=N\int_0^T ds\, \sum_{y<z}d_{yz}(s)\phi_{yz}(s)\equiv N \phi_{\rm d}.
\end{align}
Importantly, the large-$N$ scaling of the cumulants of $\Phi_{\rm d}$ are identical to the cumulants of our generalized current $J_T$ (cf.~\eqref{eq:Jd}) of interest,
\begin{equation}\label{eq:iid}
\begin{split}
&\lim_{N\to\infty}\frac{1}{N}{\rm Var}(\Phi_{\rm d})={\rm Var}(J_T) \\
&\lim_{N\to\infty}\frac{1}{N}\langle\Phi_{\rm d}\rangle=\langle J_T\rangle,
\end{split}
\end{equation}
since our ensemble of copies are independent and identically distributed.
Furthermore, they are encoded in the large deviation rate function $I(\phi_{\rm d})$ for the generalized current.
Thus, by bounding $I(\phi_{\rm d})$, as we now do, we constrain the generalized-current fluctuations.

\section{Bounding the large deviation rate function}

Remarkably, ${\mathcal I}$ in  \eqref{eq:Iinstant} has the exact same functional form as the level 2.5 large deviation rate function for long-time-averaged empirical density and currents \cite{Maes2008,Bertini2015Flows,Bertini2015Large}.
As a consequence, we can almost directly import the proof used to derive the long-time thermodynamic uncertainty relation to this situation.
As such we proceed in two steps \cite{Gingrich2016}: First, we bound ${\mathcal I}$, and then exploit the large-deviation contraction principle to obtain an inequality for the rate function $I(\phi_{\rm d})$.

As shown in \cite{Gingrich2016,Gingrich2017}, ${\mathcal I}$ satisfies a quadratic inequality, which in this situation reads
\begin{equation}
\begin{split}\label{eq:lbound}
I[&\rho(t),\phi(t)] \\
&\le\int_0^Tds\, \sum_{y<z}\frac{(\phi_{yz}(s)-j^\rho_{yz}(s))^2}{4(j^\rho_{yz}(s))^2}\sigma_{yz}^\rho(s)-S(\rho(0)||\pi),
\end{split}
\end{equation}
where $\sigma^\rho_{yz}(s)=j^\rho_{yz}(s)\ln[r_{yz}\rho_z(s)/r_{zy}\rho_y(s)]$ is the expected entropy production along jump $z\to y$ if the density were $\rho$. 

The next step is to contract down to the large deviation rate function for generalized current.
Namely, we can obtain the large deviation function for the generalized current through the minimization~\cite{Touchette2009}:
\begin{equation}\label{eq:contraction}
I(\phi_{\rm d})=\inf_{\rho(t),\phi(t)} I[\rho(t),\phi(t)],
\end{equation}
where the minimization is constrained by $\phi_{\rm d}=\int_0^Tds\, \sum_{y<z}d_{yz}(s)\phi_{yz}(s)$, the conservation of probability ${\dot \rho}_y(t)=\sum_{z\neq y}\phi_{yz}(t)$, and normalization $\sum_y\rho_y(t)=1$.
However, an upper bound to such a minimization can be obtained by choosing any pair of $\rho$ and $\phi$ consistent with the constraints.
We choose the time-independent pair
\begin{equation}
\rho_y(t)=\pi_y,\qquad \phi_{yz}(t)=\frac{\phi_{\rm d}}{\langle J_T\rangle}j^\pi_{yz}.
\end{equation}
Substituting into \eqref{eq:contraction}, while exploiting \eqref{eq:lbound}, we obtain the quadratic bound
\begin{align}
I(\phi_{\rm d})&\le \frac{(\phi_{\rm d}-\langle J_T\rangle)^2}{4\langle J_T\rangle^2}\int_0^Tds \sum_{y<z}\sigma_{yz}^\pi\\ \label{eq:quadBound}
&=\frac{(\phi_{\rm d}-\langle J_T\rangle)^2}{4\langle J_T\rangle^2}\langle \Sigma_T\rangle 
\end{align}
in terms of the time-integrated steady-state entropy production $\langle \Sigma_T\rangle =T\Sigma^\pi=T\sum_{y<z}\sigma_{yz}^\pi$.

The finite-time uncertainty relation \eqref{eq:TUR} now follows readily, by observing that the quadratic bound is zero at the typical value, $I(\langle J_T\rangle)=0$, and that the second derivative of $I(\phi_{\rm d})$ at its minimum encodes the large $N$ scaling of the variance:
\begin{equation}
\lim_{N\to\infty}\frac{1}{N}{\rm Var}(\Phi_{\rm d})=\frac{1}{I^{\prime\prime}(\langle J_T\rangle)}\ge2\langle J_T\rangle^2/\langle \Sigma_T\rangle,
\end{equation} 
by \eqref{eq:quadBound}.
Combining this inequality with the independent-identically-distributed nature of the copies \eqref{eq:iid} leads to the thermodynamic uncertainty relation in \eqref{eq:TUR}.

\section{Discussion}

Remarkably, the finite-time uncertainty relation can be derived in almost the exact same manner as the long-time uncertainty relation using a large deviation theory for an ensemble of many copies.
Consequently, this finite-time uncertainty relation is expected to also hold for diffusion processes, since the large deviation function for diffusions has a quadratic structure identical to \eqref{eq:lbound}~\cite{Gingrich2017,Maes2008Diff}.
Similarly, we expect that tighter-than-quadratic bounds~\cite{Pietzonka2016Affinity,Polettini2016} will also hold for finite times.
Extending these constructions to an uncertainty relation for finite-time first-passage-time fluctuations would be an interesting and useful extension (cf.~\cite{Gingrich2017FPT}).
However, an extension to a discrete-time process appears untenable~\cite{Shiraishi2017}.

\begin{acknowledgments}
We gratefully acknowledge the Gordon and Betty Moore Foundation for supporting TRG and JMH as Physics of Living Systems Fellows through Grant GBMF4513.
\end{acknowledgments}

\bibliography{refs.bib}

\begin{thebibliography}{27}%
\makeatletter
\providecommand \@ifxundefined [1]{%
 \@ifx{#1\undefined}
}%
\providecommand \@ifnum [1]{%
 \ifnum #1\expandafter \@firstoftwo
 \else \expandafter \@secondoftwo
 \fi
}%
\providecommand \@ifx [1]{%
 \ifx #1\expandafter \@firstoftwo
 \else \expandafter \@secondoftwo
 \fi
}%
\providecommand \natexlab [1]{#1}%
\providecommand \enquote  [1]{``#1''}%
\providecommand \bibnamefont  [1]{#1}%
\providecommand \bibfnamefont [1]{#1}%
\providecommand \citenamefont [1]{#1}%
\providecommand \href@noop [0]{\@secondoftwo}%
\providecommand \href [0]{\begingroup \@sanitize@url \@href}%
\providecommand \@href[1]{\@@startlink{#1}\@@href}%
\providecommand \@@href[1]{\endgroup#1\@@endlink}%
\providecommand \@sanitize@url [0]{\catcode `\\12\catcode `\$12\catcode
  `\&12\catcode `\#12\catcode `\^12\catcode `\_12\catcode `\%12\relax}%
\providecommand \@@startlink[1]{}%
\providecommand \@@endlink[0]{}%
\providecommand \url  [0]{\begingroup\@sanitize@url \@url }%
\providecommand \@url [1]{\endgroup\@href {#1}{\urlprefix }}%
\providecommand \urlprefix  [0]{URL }%
\providecommand \Eprint [0]{\href }%
\providecommand \doibase [0]{http://dx.doi.org/}%
\providecommand \selectlanguage [0]{\@gobble}%
\providecommand \bibinfo  [0]{\@secondoftwo}%
\providecommand \bibfield  [0]{\@secondoftwo}%
\providecommand \translation [1]{[#1]}%
\providecommand \BibitemOpen [0]{}%
\providecommand \bibitemStop [0]{}%
\providecommand \bibitemNoStop [0]{.\EOS\space}%
\providecommand \EOS [0]{\spacefactor3000\relax}%
\providecommand \BibitemShut  [1]{\csname bibitem#1\endcsname}%
\let\auto@bib@innerbib\@empty
\bibitem [{\citenamefont {Barato}\ and\ \citenamefont
  {Seifert}(2015{\natexlab{a}})}]{Barato2015}%
  \BibitemOpen
  \bibfield  {author} {\bibinfo {author} {\bibfnamefont {A.~C.}\ \bibnamefont
  {Barato}}\ and\ \bibinfo {author} {\bibfnamefont {U.}~\bibnamefont
  {Seifert}},\ }\href {\doibase 10.1103/PhysRevLett.114.158101} {\bibfield
  {journal} {\bibinfo  {journal} {Phys. Rev. Lett.}\ }\textbf {\bibinfo
  {volume} {114}},\ \bibinfo {pages} {158101} (\bibinfo {year}
  {2015}{\natexlab{a}})}\BibitemShut {NoStop}%
\bibitem [{\citenamefont {Pietzonka}\ \emph
  {et~al.}(2016{\natexlab{a}})\citenamefont {Pietzonka}, \citenamefont
  {Barato},\ and\ \citenamefont {Seifert}}]{Pietzonka2016Universal}%
  \BibitemOpen
  \bibfield  {author} {\bibinfo {author} {\bibfnamefont {P.}~\bibnamefont
  {Pietzonka}}, \bibinfo {author} {\bibfnamefont {A.~C.}\ \bibnamefont
  {Barato}}, \ and\ \bibinfo {author} {\bibfnamefont {U.}~\bibnamefont
  {Seifert}},\ }\href {\doibase 10.1103/PhysRevE.93.052145} {\bibfield
  {journal} {\bibinfo  {journal} {Phys. Rev. E}\ }\textbf {\bibinfo {volume}
  {93}},\ \bibinfo {pages} {052145} (\bibinfo {year}
  {2016}{\natexlab{a}})}\BibitemShut {NoStop}%
\bibitem [{\citenamefont {Gingrich}\ \emph {et~al.}(2016)\citenamefont
  {Gingrich}, \citenamefont {Horowitz}, \citenamefont {Perunov},\ and\
  \citenamefont {England}}]{Gingrich2016}%
  \BibitemOpen
  \bibfield  {author} {\bibinfo {author} {\bibfnamefont {T.~R.}\ \bibnamefont
  {Gingrich}}, \bibinfo {author} {\bibfnamefont {J.~M.}\ \bibnamefont
  {Horowitz}}, \bibinfo {author} {\bibfnamefont {N.}~\bibnamefont {Perunov}}, \
  and\ \bibinfo {author} {\bibfnamefont {J.~L.}\ \bibnamefont {England}},\
  }\href {\doibase 10.1103/PhysRevLett.116.120601} {\bibfield  {journal}
  {\bibinfo  {journal} {Phys. Rev. Lett.}\ }\textbf {\bibinfo {volume} {116}},\
  \bibinfo {pages} {120601} (\bibinfo {year} {2016})}\BibitemShut {NoStop}%
\bibitem [{\citenamefont {Gingrich}\ \emph {et~al.}(2017)\citenamefont
  {Gingrich}, \citenamefont {Rotskoff},\ and\ \citenamefont
  {Horowitz}}]{Gingrich2017}%
  \BibitemOpen
  \bibfield  {author} {\bibinfo {author} {\bibfnamefont {T.~R.}\ \bibnamefont
  {Gingrich}}, \bibinfo {author} {\bibfnamefont {G.~M.}\ \bibnamefont
  {Rotskoff}}, \ and\ \bibinfo {author} {\bibfnamefont {J.~M.}\ \bibnamefont
  {Horowitz}},\ }\href {\doibase 10.1088/1751-8121/aa672f} {\bibfield
  {journal} {\bibinfo  {journal} {J. Phys. A: Math. Theor.}\ }\textbf {\bibinfo
  {volume} {50}},\ \bibinfo {pages} {184004} (\bibinfo {year}
  {2017})}\BibitemShut {NoStop}%
\bibitem [{\citenamefont {Polettini}\ \emph {et~al.}(2016)\citenamefont
  {Polettini}, \citenamefont {Lazarescu},\ and\ \citenamefont
  {Esposito}}]{Polettini2016}%
  \BibitemOpen
  \bibfield  {author} {\bibinfo {author} {\bibfnamefont {M.}~\bibnamefont
  {Polettini}}, \bibinfo {author} {\bibfnamefont {A.}~\bibnamefont
  {Lazarescu}}, \ and\ \bibinfo {author} {\bibfnamefont {M.}~\bibnamefont
  {Esposito}},\ }\href {\doibase 10.1103/PhysRevE.94.052104} {\bibfield
  {journal} {\bibinfo  {journal} {Phys. Rev. E}\ }\textbf {\bibinfo {volume}
  {94}},\ \bibinfo {pages} {052104} (\bibinfo {year} {2016})}\BibitemShut
  {NoStop}%
\bibitem [{\citenamefont {Maes}(2017)}]{Maes2017}%
  \BibitemOpen
  \bibfield  {author} {\bibinfo {author} {\bibfnamefont {C.}~\bibnamefont
  {Maes}},\ }\href@noop {} {\enquote {\bibinfo {title} {Frenetic bounds on the
  entropy production},}\ } (\bibinfo {year} {2017}),\ \bibinfo {note}
  {arXiv:1705.07412}\BibitemShut {NoStop}%
\bibitem [{\citenamefont {Pietzonka}\ \emph
  {et~al.}(2016{\natexlab{b}})\citenamefont {Pietzonka}, \citenamefont
  {Barato},\ and\ \citenamefont {Seifert}}]{Pietzonka2016Affinity}%
  \BibitemOpen
  \bibfield  {author} {\bibinfo {author} {\bibfnamefont {P.}~\bibnamefont
  {Pietzonka}}, \bibinfo {author} {\bibfnamefont {A.~C.}\ \bibnamefont
  {Barato}}, \ and\ \bibinfo {author} {\bibfnamefont {U.}~\bibnamefont
  {Seifert}},\ }\href {\doibase 10.1088/1751-8113/49/34/34LT01} {\bibfield
  {journal} {\bibinfo  {journal} {J. Phys. A: Math. Theor.}\ }\textbf {\bibinfo
  {volume} {49}},\ \bibinfo {pages} {34LT01} (\bibinfo {year}
  {2016}{\natexlab{b}})}\BibitemShut {NoStop}%
\bibitem [{\citenamefont {Barato}\ and\ \citenamefont
  {Seifer}(2015)}]{Barato2015Fano}%
  \BibitemOpen
  \bibfield  {author} {\bibinfo {author} {\bibfnamefont {A.~C.}\ \bibnamefont
  {Barato}}\ and\ \bibinfo {author} {\bibfnamefont {U.}~\bibnamefont
  {Seifer}},\ }\href@noop {} {\bibfield  {journal} {\bibinfo  {journal} {J.
  Phys. Chem. B}\ }\textbf {\bibinfo {volume} {119}},\ \bibinfo {pages} {6555}
  (\bibinfo {year} {2015})}\BibitemShut {NoStop}%
\bibitem [{\citenamefont {Pietzonka}\ \emph
  {et~al.}(2016{\natexlab{c}})\citenamefont {Pietzonka}, \citenamefont
  {Barato},\ and\ \citenamefont {Seifert}}]{Pietzonka2016MolecularMotors}%
  \BibitemOpen
  \bibfield  {author} {\bibinfo {author} {\bibfnamefont {P.}~\bibnamefont
  {Pietzonka}}, \bibinfo {author} {\bibfnamefont {A.~C.}\ \bibnamefont
  {Barato}}, \ and\ \bibinfo {author} {\bibfnamefont {U.}~\bibnamefont
  {Seifert}},\ }\href {\doibase 10.1088/1742-5468/2016/12/124004} {\bibfield
  {journal} {\bibinfo  {journal} {J. Stat. Mech. Theor. Exp.}\ }\textbf
  {\bibinfo {volume} {12}},\ \bibinfo {pages} {124004} (\bibinfo {year}
  {2016}{\natexlab{c}})}\BibitemShut {NoStop}%
\bibitem [{\citenamefont {Pietzonka}\ and\ \citenamefont
  {Seifert}(2017)}]{Pietzonka2017universal}%
  \BibitemOpen
  \bibfield  {author} {\bibinfo {author} {\bibfnamefont {P.}~\bibnamefont
  {Pietzonka}}\ and\ \bibinfo {author} {\bibfnamefont {U.}~\bibnamefont
  {Seifert}},\ }\href@noop {} {\enquote {\bibinfo {title} {Universal trade-off
  between power, efficiency and constancy in steady-state heat engines},}\ }
  (\bibinfo {year} {2017}),\ \bibinfo {note} {arXiv:1705.05817}\BibitemShut
  {NoStop}%
\bibitem [{\citenamefont {Barato}\ and\ \citenamefont
  {Seifert}(2015{\natexlab{b}})}]{Barat2015Dispersion}%
  \BibitemOpen
  \bibfield  {author} {\bibinfo {author} {\bibfnamefont {A.~C.}\ \bibnamefont
  {Barato}}\ and\ \bibinfo {author} {\bibfnamefont {U.}~\bibnamefont
  {Seifert}},\ }\href@noop {} {\bibfield  {journal} {\bibinfo  {journal} {Phys.
  Rev. E}\ }\textbf {\bibinfo {volume} {92}},\ \bibinfo {pages} {032127}
  (\bibinfo {year} {2015}{\natexlab{b}})}\BibitemShut {NoStop}%
\bibitem [{\citenamefont {Hyeon}\ and\ \citenamefont
  {Hwang}(2017)}]{Hyeon2017}%
  \BibitemOpen
  \bibfield  {author} {\bibinfo {author} {\bibfnamefont {C.}~\bibnamefont
  {Hyeon}}\ and\ \bibinfo {author} {\bibfnamefont {W.}~\bibnamefont {Hwang}},\
  }\href {\doibase 10.1103/PhysRevE.96.012156} {\bibfield  {journal} {\bibinfo
  {journal} {Phys. Rev. E}\ }\textbf {\bibinfo {volume} {96}},\ \bibinfo
  {pages} {012156} (\bibinfo {year} {2017})}\BibitemShut {NoStop}%
\bibitem [{\citenamefont {Nguyen}\ and\ \citenamefont
  {Vaikuntanathan}(2016)}]{Nguyen2016}%
  \BibitemOpen
  \bibfield  {author} {\bibinfo {author} {\bibfnamefont {M.}~\bibnamefont
  {Nguyen}}\ and\ \bibinfo {author} {\bibfnamefont {S.}~\bibnamefont
  {Vaikuntanathan}},\ }\href@noop {} {\bibfield  {journal} {\bibinfo  {journal}
  {PNAS}\ }\textbf {\bibinfo {volume} {113}},\ \bibinfo {pages} {14231}
  (\bibinfo {year} {2016})}\BibitemShut {NoStop}%
\bibitem [{\citenamefont {Falasco}\ \emph {et~al.}(2016)\citenamefont
  {Falasco}, \citenamefont {Pfaller}, \citenamefont {Bregulla}, \citenamefont
  {Cichos},\ and\ \citenamefont {Kroy}}]{Falasco2016}%
  \BibitemOpen
  \bibfield  {author} {\bibinfo {author} {\bibfnamefont {G.}~\bibnamefont
  {Falasco}}, \bibinfo {author} {\bibfnamefont {R.}~\bibnamefont {Pfaller}},
  \bibinfo {author} {\bibfnamefont {A.~P.}\ \bibnamefont {Bregulla}}, \bibinfo
  {author} {\bibfnamefont {F.}~\bibnamefont {Cichos}}, \ and\ \bibinfo {author}
  {\bibfnamefont {K.}~\bibnamefont {Kroy}},\ }\href@noop {} {\bibfield
  {journal} {\bibinfo  {journal} {Phys. Rev. E}\ }\textbf {\bibinfo {volume}
  {94}},\ \bibinfo {pages} {030602(R)} (\bibinfo {year} {2016})}\BibitemShut
  {NoStop}%
\bibitem [{\citenamefont {Guioth}\ and\ \citenamefont
  {Lacoste}(2016)}]{Guioth2016}%
  \BibitemOpen
  \bibfield  {author} {\bibinfo {author} {\bibfnamefont {J.}~\bibnamefont
  {Guioth}}\ and\ \bibinfo {author} {\bibfnamefont {D.}~\bibnamefont
  {Lacoste}},\ }\href@noop {} {\bibfield  {journal} {\bibinfo  {journal} {EPL
  (Europhysics Letters)}\ }\textbf {\bibinfo {volume} {115}},\ \bibinfo {pages}
  {60007} (\bibinfo {year} {2016})}\BibitemShut {NoStop}%
\bibitem [{\citenamefont {Garrahan}(2017)}]{Garrahan2017}%
  \BibitemOpen
  \bibfield  {author} {\bibinfo {author} {\bibfnamefont {J.~P.}\ \bibnamefont
  {Garrahan}},\ }\href {\doibase 10.1103/PhysRevE.95.032134} {\bibfield
  {journal} {\bibinfo  {journal} {Physical Review E}\ }\textbf {\bibinfo
  {volume} {95}},\ \bibinfo {pages} {032134} (\bibinfo {year}
  {2017})}\BibitemShut {NoStop}%
\bibitem [{\citenamefont {Gingrich}\ and\ \citenamefont
  {Horowitz}(2017)}]{Gingrich2017FPT}%
  \BibitemOpen
  \bibfield  {author} {\bibinfo {author} {\bibfnamefont {T.~R.}\ \bibnamefont
  {Gingrich}}\ and\ \bibinfo {author} {\bibfnamefont {J.~M.}\ \bibnamefont
  {Horowitz}},\ }\href@noop {} {\enquote {\bibinfo {title} {Fundamental bounds
  on first passage time fluctuations for currents},}\ } (\bibinfo {year}
  {2017}),\ \bibinfo {note} {arXiv:1706.09027}\BibitemShut {NoStop}%
\bibitem [{\citenamefont {Touchette}(2009)}]{Touchette2009}%
  \BibitemOpen
  \bibfield  {author} {\bibinfo {author} {\bibfnamefont {H.}~\bibnamefont
  {Touchette}},\ }\href {\doibase 10.1016/j.physrep.2009.05.002} {\bibfield
  {journal} {\bibinfo  {journal} {Physics Reports}\ }\textbf {\bibinfo {volume}
  {478}},\ \bibinfo {pages} {1} (\bibinfo {year} {2009})}\BibitemShut {NoStop}%
\bibitem [{\citenamefont {Pietzonka}\ \emph {et~al.}(2017)\citenamefont
  {Pietzonka}, \citenamefont {Ritort},\ and\ \citenamefont
  {Seifert}}]{Pietzonka2017}%
  \BibitemOpen
  \bibfield  {author} {\bibinfo {author} {\bibfnamefont {P.}~\bibnamefont
  {Pietzonka}}, \bibinfo {author} {\bibfnamefont {F.}~\bibnamefont {Ritort}}, \
  and\ \bibinfo {author} {\bibfnamefont {U.}~\bibnamefont {Seifert}},\
  }\href@noop {} {\bibfield  {journal} {\bibinfo  {journal} {Phys. Rev. E}\
  }\textbf {\bibinfo {volume} {96}},\ \bibinfo {pages} {012101} (\bibinfo
  {year} {2017})}\BibitemShut {NoStop}%
\bibitem [{\citenamefont {Pigolotti}\ \emph {et~al.}(2017)\citenamefont
  {Pigolotti}, \citenamefont {Neri}, \citenamefont {Rold{\'a}n},\ and\
  \citenamefont {J{\"u}licher}}]{Pigolotti2017}%
  \BibitemOpen
  \bibfield  {author} {\bibinfo {author} {\bibfnamefont {S.}~\bibnamefont
  {Pigolotti}}, \bibinfo {author} {\bibfnamefont {I.}~\bibnamefont {Neri}},
  \bibinfo {author} {\bibfnamefont {{\'E}.}~\bibnamefont {Rold{\'a}n}}, \ and\
  \bibinfo {author} {\bibfnamefont {F.}~\bibnamefont {J{\"u}licher}},\
  }\href@noop {} {\enquote {\bibinfo {title} {Generic properties of stochastic
  entropy production},}\ } (\bibinfo {year} {2017}),\ \bibinfo {note}
  {arXiv:1704.04061}\BibitemShut {NoStop}%
\bibitem [{\citenamefont {Seifert}(2012)}]{Seifert2012}%
  \BibitemOpen
  \bibfield  {author} {\bibinfo {author} {\bibfnamefont {U.}~\bibnamefont
  {Seifert}},\ }\href {\doibase 10.1088/0034-4885/75/12/126001} {\bibfield
  {journal} {\bibinfo  {journal} {Rep. Prog. Phys.}\ }\textbf {\bibinfo
  {volume} {75}},\ \bibinfo {pages} {126001} (\bibinfo {year}
  {2012})}\BibitemShut {NoStop}%
\bibitem [{\citenamefont {Maes}\ \emph
  {et~al.}(2008{\natexlab{a}})\citenamefont {Maes}, \citenamefont {Netocny},\
  and\ \citenamefont {Wynants}}]{Maes2008On}%
  \BibitemOpen
  \bibfield  {author} {\bibinfo {author} {\bibfnamefont {C.}~\bibnamefont
  {Maes}}, \bibinfo {author} {\bibfnamefont {K.}~\bibnamefont {Netocny}}, \
  and\ \bibinfo {author} {\bibfnamefont {B.}~\bibnamefont {Wynants}},\
  }\href@noop {} {\bibfield  {journal} {\bibinfo  {journal} {Markov Process.
  Relat.}\ }\textbf {\bibinfo {volume} {14}},\ \bibinfo {pages} {445} (\bibinfo
  {year} {2008}{\natexlab{a}})}\BibitemShut {NoStop}%
\bibitem [{\citenamefont {Bertini}\ \emph
  {et~al.}(2015{\natexlab{a}})\citenamefont {Bertini}, \citenamefont
  {Faggionato},\ and\ \citenamefont {Gabrielli}}]{Bertini2015Flows}%
  \BibitemOpen
  \bibfield  {author} {\bibinfo {author} {\bibfnamefont {L.}~\bibnamefont
  {Bertini}}, \bibinfo {author} {\bibfnamefont {A.}~\bibnamefont {Faggionato}},
  \ and\ \bibinfo {author} {\bibfnamefont {D.}~\bibnamefont {Gabrielli}},\
  }\href {\doibase 10.1016/j.spa.2015.02.001} {\bibfield  {journal} {\bibinfo
  {journal} {Stoch. Proc. Appl.}\ }\textbf {\bibinfo {volume} {125}},\ \bibinfo
  {pages} {2786} (\bibinfo {year} {2015}{\natexlab{a}})}\BibitemShut {NoStop}%
\bibitem [{\citenamefont {Bertini}\ \emph
  {et~al.}(2015{\natexlab{b}})\citenamefont {Bertini}, \citenamefont
  {Faggionato},\ and\ \citenamefont {Gabrielli}}]{Bertini2015Large}%
  \BibitemOpen
  \bibfield  {author} {\bibinfo {author} {\bibfnamefont {L.}~\bibnamefont
  {Bertini}}, \bibinfo {author} {\bibfnamefont {A.}~\bibnamefont {Faggionato}},
  \ and\ \bibinfo {author} {\bibfnamefont {D.}~\bibnamefont {Gabrielli}},\
  }\href {\doibase 10.1214/14-AIHP601} {\bibfield  {journal} {\bibinfo
  {journal} {Ann. I. H. Poincare-PR}\ }\textbf {\bibinfo {volume} {51}},\
  \bibinfo {pages} {867} (\bibinfo {year} {2015}{\natexlab{b}})}\BibitemShut
  {NoStop}%
\bibitem [{\citenamefont {Maes}\ and\ \citenamefont
  {Neto{\v{c}}n{\`y}}(2008)}]{Maes2008}%
  \BibitemOpen
  \bibfield  {author} {\bibinfo {author} {\bibfnamefont {C.}~\bibnamefont
  {Maes}}\ and\ \bibinfo {author} {\bibfnamefont {K.}~\bibnamefont
  {Neto{\v{c}}n{\`y}}},\ }\href {\doibase 10.1209/0295-5075/82/30003}
  {\bibfield  {journal} {\bibinfo  {journal} {EPL (Europhysics Letters)}\
  }\textbf {\bibinfo {volume} {82}},\ \bibinfo {pages} {30003} (\bibinfo {year}
  {2008})}\BibitemShut {NoStop}%
\bibitem [{\citenamefont {Maes}\ \emph
  {et~al.}(2008{\natexlab{b}})\citenamefont {Maes}, \citenamefont
  {Neto{\v{c}}n{\`y}},\ and\ \citenamefont {Wynants}}]{Maes2008Diff}%
  \BibitemOpen
  \bibfield  {author} {\bibinfo {author} {\bibfnamefont {C.}~\bibnamefont
  {Maes}}, \bibinfo {author} {\bibfnamefont {K.}~\bibnamefont
  {Neto{\v{c}}n{\`y}}}, \ and\ \bibinfo {author} {\bibfnamefont
  {B.}~\bibnamefont {Wynants}},\ }\href@noop {} {\bibfield  {journal} {\bibinfo
   {journal} {Physica A}\ }\textbf {\bibinfo {volume} {387}},\ \bibinfo {pages}
  {2675} (\bibinfo {year} {2008}{\natexlab{b}})}\BibitemShut {NoStop}%
\bibitem [{\citenamefont {Shiraishi}(2017)}]{Shiraishi2017}%
  \BibitemOpen
  \bibfield  {author} {\bibinfo {author} {\bibfnamefont {N.}~\bibnamefont
  {Shiraishi}},\ }\href@noop {} {\enquote {\bibinfo {title} {Finite-time
  thermodynamic uncerainty relation do not hold for discrete-time {M}arkov
  process},}\ } (\bibinfo {year} {2017}),\ \bibinfo {note}
  {arXiv:1706.00892}\BibitemShut {NoStop}%
\end{thebibliography}%

\end{document}